\documentclass[a4paper]{article}

\usepackage{amsmath}
\usepackage{amssymb}
\usepackage{color}
\usepackage{graphicx}
\usepackage{authblk}

\begin{document}

\title{The origin of the power--law form \\
 of the extragalactic gamma--ray flux}

\author[1]{\normalsize Paolo Lipari}

\affil[1]{\footnotesize INFN, Sezione Roma ``Sapienza'',
Piazzale Aldo Moro 2, 00185 Roma, Italy}

\date{\small January 3, 2020}


\maketitle

\begin{abstract}
The Fermi--LAT telescope has measured the extragalactic
gamma--ray background (EGB)
generated by the ensemble of all extragalactic sources. The energy
distribution of the EGB is well described as a power--law with
a spectral index approximately equal to 2.3, and an exponential cutoff,
that is consistent with being the effect of absorption
of high energy photons assuming an emission that is an
unbroken power--law spectrum.
The observations of the Fermi telescope have also resolved the EGB,
determining that most of it is formed by the sum of the contributions
of Active Galactic Nuclei (AGN) of the blazar class. The
spectra of the individual AGN sources have a broad
range of spectral shapes, and the brightest and most precisely
measured sources have curved spectra that can be
fitted with the ``log--parabola'' (or log--normal) form.
It might appear surprising that the sum of contributions
with different spectral shapes generate a spectrum that has a
simple power--law form.
We suggest that this fact reveals some important properties
for the ensemble of the extragalactic sources,
and indicates that the blazar high energy emission 
can be considered as a ``critical phenomenon''.
A curved log--parabola (log--normal) form is also required to fit
the spectra of many Galactic gamma--ray sources, including the emission
from the brightest supernova remnants (SNR), and an intriguing possibility is that also the
Galactic cosmic rays are injected in interstellar space by their sources
not with a single universal power--law spectrum, as predicted by the
commonly accepted models, but instead with a broad range of spectral shapes. 
\end{abstract}

\section{Introduction} 
\label{sec:intro}
The Fermi--LAT telescope has measured \cite{Ackermann:2014usa}
the extragalactic gamma--ray flux,
generated by the ensemble of all extragalactic sources,
in the energy range 0.1--820~GeV.
The total flux (commonly called Extragalactic Gamma Background or EGB)
can be decomposed into a component due to the ensemble of resolved
extragalactic point sources, and a second component (the Isotropic
Gamma Ray Background or IGRB) that accounts for all other 
emissions, including unresolved faint point sources.
This decomposition obviously depends on the instrument
sensitivity and integration time.

Both the EGB and the IGRB are in good approximation isotropic
(reflecting the homogeneity and isotropy of the universe), and
their energy spectra have been fitted
in \cite{Ackermann:2014usa} with the functional form:
\begin{equation}
\phi_\gamma (E) = K_\gamma ~E^{-\alpha} ~e^{-E/E_{\rm cut}}
\label{eq:fit_form}
\end{equation}
that is a power--law with an exponential cutoff.

The Fermi--LAT Collaboration has presented
three estimates for the EGB and IGRB
obtained using different models for the Galactic foreground,
these estimates are however close to each other with differences
that are negligible for the purposes of our work.
Averaging the results of the three fits one obtains:
$\alpha = 2.30\pm 0.02$, $E_{\rm cut} = 330\pm 70$~GeV for the EGB, 
and 
$\alpha = 2.29\pm 0.02$, $E_{\rm cut} = 239 \pm 50$~GeV for the IGRB.
One can note that the spectral indices
for the EGB and the IGRB are consistent with being equal,
suggesting a common origin.
It is also likely that, correcting for absorption effects,
the spectra are reasonably well
described by a simple power--law form in the entire energy range of
the Fermi--LAT observations. This is because 
the observed spectral cutoffs are consistent with being
the distortions generated
by the absorption of high energy gamma--rays during propagation,
assuming an emission that is an unbroken power--law.
The main source of absorption is due to pair production interactions
($\gamma \gamma \to e^-e^+$) with the target photons that form
the intergalactic radiation fields. 
The cutoff energy for the IGRB is smaller than the cutoff for the
EGB, but this can be naturally explained assuming that a large fraction
of the unresolved flux is due to faint, distant sources
that are more absorbed.

The Fermi--LAT observations also show that approximately
one half of the total (EGB) extragalactic flux is due to 
an ensemble of point sources, most of them
Active Galactic Nuclei (AGN) of the blazar class.
The observations of the point sources allow
to model their luminosity function and cosmological evolution
and then to estimate the flux 
of those that are too faint to be resolved.
These studies indicate that blazars also account for most of the
IGRB \cite{TheFermi-LAT:2015ykq}.
Other sources (such as normal
or starburst Galaxies or Dark Matter self--annihilation)
can contribute only a fraction of order 10\% or less.

Some questions emerge immediately from these results.
What are the astrophysical mechanisms
that generate the extragalactic gamma--ray emission?
What is the origin of the power--law form of the spectrum?
Why the spectral index has the value $\alpha \simeq 2.30$~?

Simple considerations suggest that if a spectrum
has a power--law form with slope $\alpha$,
and is formed by the sum of distinct components,
then also the energy distributions of the components
have a power--law form with the same spectral index.
This is because it appears difficult to combine
spectra of different shapes to form a sum
that has a featureless power--law form.
Also in the case where the individual components are all
of power--law form, but have different slopes,
the spectrum of the sum is not a simple power--law, 
but hardens gradually, with an energy dependent slope.

On the other hand, the spectra of the extragalactic point sources
resolved by Fermi--LAT have a broad range of spectral shapes,
that in many cases are not simple power--laws, but are ``curved''
(in a log--log representation) with an energy dependent spectral index.
Nonetheless, the sum of these contributions generate a total flux
of simple power--law form.
This surprising result indicates that the slope
$\alpha \approx 2.30$ of the average extragalactic gamma--ray emission
is related to the properties of the {\em ensemble of the sources}
and is not a ``universal slope'' that describes the emission
of the individual sources.

In this work we address this problem and investigate the origin of the
spectral index of the extragalactic gamma--ray flux.
The paper is organized as follows.
In the next section we study the spectra of the extragalactic
sources observed by Fermi--LAT, and discuss the properties
of the log--parabola form that is used to fit 18\% of the
sources (but accounts for over 60\% of the total resolved extragalactic
flux).
In Sec.~\ref{sec:combined-spectra} we discuss the
shape of a spectrum formed by components that have different
energy distributions, and show how an ensemble of log--parabola
components can combine to generate an average that has power--law form.
In Sec.~\ref{sec:statistical} we show how the conditions
required to form a power--law spectrum from log--parabola
components can be satisfied if the luminosity of the sources
has a power--law dependence of the hardness of their spectra.
In this case the spectral index of the extragalactic flux is related
to the exponent that describes the luminosity--hardness relation.
Sec.~\ref{sec:power-laws} discusses how 
blazar emission can be described as a ``critical phenomenon''.
Section~\ref{sec:galactic} briefly comments on the cosmic ray Galactic sources
and of the possible relevance of the concepts developed here also
for Galactic accelerators. It is in fact puzzling that
most of the Supernova remnants observed by Fermi--LAT
have log--parabola spectra (that account for more than 90\% of the total
flux for this class of objects).
Sec.~\ref{sec:conclusions} contains some concluding remarks. 
An appendix very briefly discusses how simple modifications
of the standard Fermi acceleration mechanism can result in
gradually softening spectra that are in good approximation
of log--parabola form.

\section{Gamma--ray point sources}
\label{sec:point-sources}

\subsection{Spectral shapes} 
The Fermi--LAT telescope has measured the spectra of a large number
of point--like and quasi--point like sources.
The recently released fourth source catalog (4FGL)
\cite{Fermi-LAT:2019yla} lists 5066 sources
with detection significance of more than 4 sigma, and 
for each source provides a description of the spectral properties
in the form of an analytic fit. These results are of extraordinary value
to develop an understanding of the astrophysical acceleration mechanisms.

The 4FGL catalog uses three functional forms to
fit the spectra of the sources: ``Power--law'', ``Log--parabola''
and ``Cutoff''. The ``Cutoff'' form
(a power--law with a super--exponential cutoff)
is used in the 4FGL to fit the spectra of 220 sources
(218 Pulsars, the Small Magellanic Cloud, and the blazar 3C 454.3).
In the present work we are not interested in Pulsars,
and this spectral form will not be discussed further.

Most of the sources in the catalog (3543)
are fitted with the 2--parameter Power--Law form:
\begin{equation}
\phi_\gamma (E)= \phi_0~\left (\frac{E}{E_0} \right )^{-\alpha} ~.
\label{eq:power_form}
\end{equation}
In this expression $E_0$ is not a parameter but
a reference energy (called the ``pivot energy'')
chosen as the energy where the error on the absolute flux is minimum.

Approximately 25\% of the sources in the catalog (1303) 
are fitted with the 3--parameter ``log--parabola'' form
\begin{equation}
\phi(E) = \phi_0 \; \left ( \frac{E}{E_0} \right )^{-(\alpha_0 + \beta \, \ln E/E_0)} ~.
\label{eq:log_parabola_form}
\end{equation}
As in the previous case $E_0$ is a source dependent pivot energy,
$\phi_0$ and $\alpha_0$ are the flux and
the spectral index at $E_0$, while $\beta$ gives the curvature of the spectrum.
For all sources in the catalog $\beta$ is positive,
and this corresponds to a gradually softening spectrum.

The name of this spectral shape expresses the fact
that in a log--log representation ($\log \phi(E)$ versus $\log E$)
the spectrum has the form of a parabola. This parabolic form
is conserved also when the spectrum is represented in the form
 ($\log E^n \phi(E) $ versus $\log E$) for any value of the exponent $n$.

The energy dependent slope $\alpha (E)$ of a log--parabola spectrum is: 
\begin{equation}
 \alpha(E)
 = -\frac{d\log \phi(E)} {d\log E}
 = \alpha_0 + 2 \beta \, \ln \frac{E}{E_0}
 \label{eq:alpha_logparabola1}
\end{equation}
and grows linearly with $\log E$ with coefficient $2 \beta$
taking all real values, from $-\infty$ at very low energy to $+\infty$ at very high energy.

The log--parabola expression can also be rewritten in the form of a log--normal distribution as: 
\begin{equation}
\phi(E)
= \phi_\dagger \; \left ( \frac{E}{E_\dagger } \right )^{-\beta \, \ln E/E_\dagger} 
= \phi_\dagger
~e^{-\beta \; (\ln E -\ln E_\dagger)^2} 
\label{eq:log_parabola_form2}
\end{equation}
In this expression we have eliminated the arbitrary reference energy $E_0$,
and introduced as a new parameter
$E_\dagger$ the energy where the flux has its maximum
(and where the spectral index vanishes: $\alpha (E_\dagger)=0$);
$\phi_\dagger$ is the value of the flux at $E = E_\dagger$. 
The new parameters $E_\dagger$ and $\phi_\dagger$
can be obtained from $E_0$ and $\alpha_0$ as: 
\begin{equation}
E_\dagger = E_0 \; e^{-\alpha_0/(2 \, \beta)}
\end{equation}
\begin{equation}
 \phi_\dagger = \phi_0 \; e^{\alpha_0^2/(4 \, \beta)} ~.
\end{equation}

The value $E_\dagger$ is typically (for $\beta$ small)
much below the energy range where the spectra are measured,
and for this reason we find it more convenient
to parametrize the log--parabola form in terms of a different
quantity, the characteristic energy $E_*$ defined as
the energy where the Spectral Energy Distributions (SED)
$S(E) = E^2 \, \phi(E)$ has its maximum,
or equivalently where the spectral index takes the value $\alpha(E_*) = 2$.
The SED of a log--parabola spectrum is symmetric around the characteristic
energy $E_*$, so that the energy fluxes integrated
below and above $E_*$ are equal.
The log--parabola expression can be written in terms of $E_*$ in the form:
\begin{equation}
\phi(E)
= \phi_* \; \left ( \frac{E}{E_* } \right )^{-(2 +\beta \, \ln E/E_*)} 
= \phi_*
~e^{-(2 \, \ln E/E_* + \beta \, \ln^2 E/E_*)} 
\label{eq:log_parabola_form1}
\end{equation}
(where $\phi_*$ is the flux at $E = E_*$).
The parameters $E_*$ and $\phi_*$ can be calculated as:
\begin{equation}
 E_*
 = E_0 \; e^{(2-\alpha_0)/(2 \, \beta)}
 = E_\dagger \; e^{1/ \beta}
\end{equation}
\begin{equation}
 \phi_* = \phi_0 \; e^{(\alpha_0^2-4)/(4 \, \beta)} 
 = \phi_\dagger \; e^{-1/ \beta} ~.
\end{equation}
The spectral index $\alpha(E)$ of the log--parabola form
can be written using the parameter $E_*$ (or $E_\dagger$) as:
\begin{equation}
 \alpha(E)
 = 2 + 2 \, \beta \, \ln \frac{E}{E_*} 
 = 2 \, \beta \, \ln \frac{E}{E_\dagger} ~. 
 \label{eq:alpha_logparabola}
\end{equation}

It is important for the following discussion to note 
that a log--parabola spectrum
(in contrast to a featureless, scale free power--law)
determines an energy scale
(that can be chosen has the energy $E_*$ where the SED has its maximum).
For a fixed value of the curvature parameter $\beta$
the shape of the spectrum then depends only on the ratio $E/E_*$.

It is also useful to note that (for $\beta >0$)
the log--parabola expression can in principle be extended
to all energies $E> 0$.
This is not possible for a power--law spectrum,
because it would result in divergences
in the number of particles and their total energy.
On the other hand for a log--parabola (log--normal) 
spectrum (when $\beta > 0$) the momenta of arbitrary order $m$
are always finite:
\begin{equation}
 \left \langle \phi(E) \; E^m \right \rangle =
 \int_0^\infty dE~\phi(E)~E^m = \phi_* \; E_*^{m+1} ~\sqrt{\frac{\pi}{\beta}}
 ~e^{(m-1)^2/(4 \, \beta)} ~.
\label{eq:moment-m}
\end{equation}
This implies can the average energy of particles in a log--parabola
spectrum is proportional to the characteristic energy:
$\langle E \rangle = E_* \; e^{-1/(4 \beta)}$. 

\subsection{Extragalactic point sources} 
\begin{figure}[bt]
\begin{center}
\includegraphics[width=15.0cm]{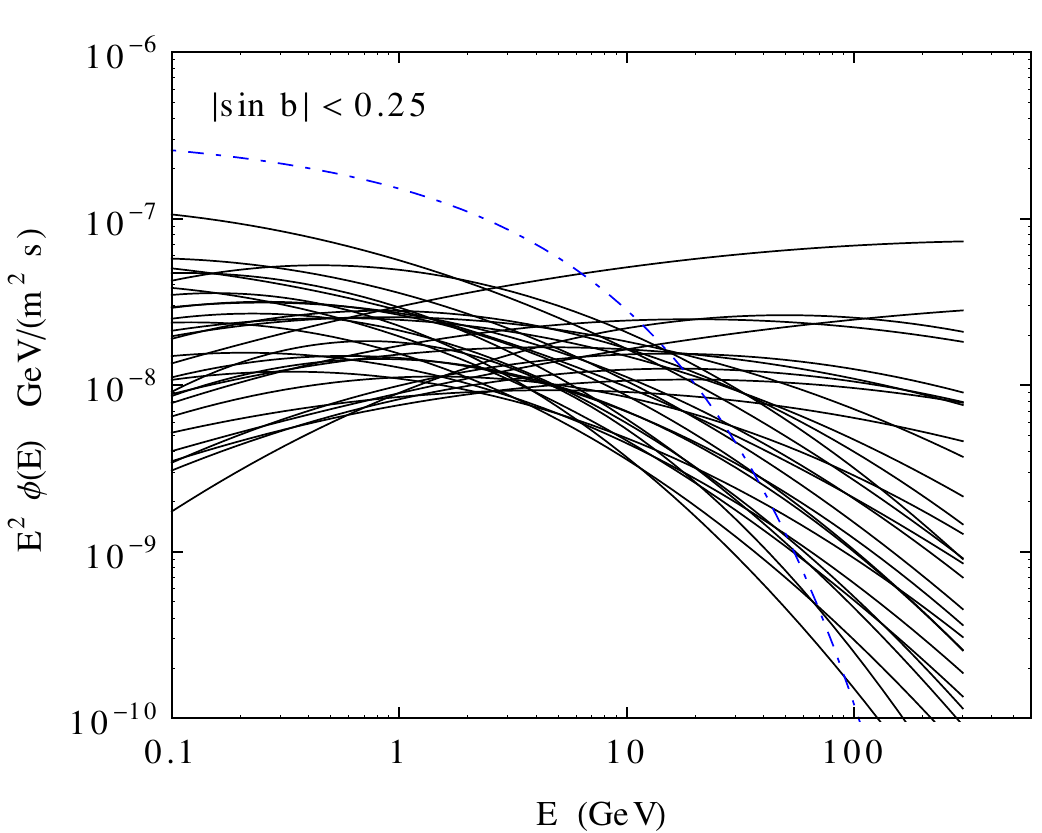}
\end{center}
\caption {\footnotesize
 Spectra of the fits to the 30 brightest extragalactic sources 
 found in the sky region $|\sin b | > 0.25$ in the 4FGL catalog.
 The dot--dashed (blue) line is the fit to the blazar 3C 454.3 (fitted
 as a power--law with a super exponential cutoff).
 The other lines are fits with the log--parabola form.
 \label{fig:gamma_resolved}}
\end{figure}

To select a sample of extragalactic sources we have
chosen the sky region $|\sin b| > 0.25$ around the two Galactic poles. 
In this part of the sky,
after the exclusion of a few objects that are classified as Pulsars
(and are therefore Galactic), the 4FGL catalog contains 3223 sources,
with a neglible contamination of Galactic objects.
The best fits to the spectra of the 30 brightest
objects in this selection are shown in Fig~\ref{fig:gamma_resolved}.

\begin{figure}[bt]
\begin{center}
\includegraphics[width=15.0cm]{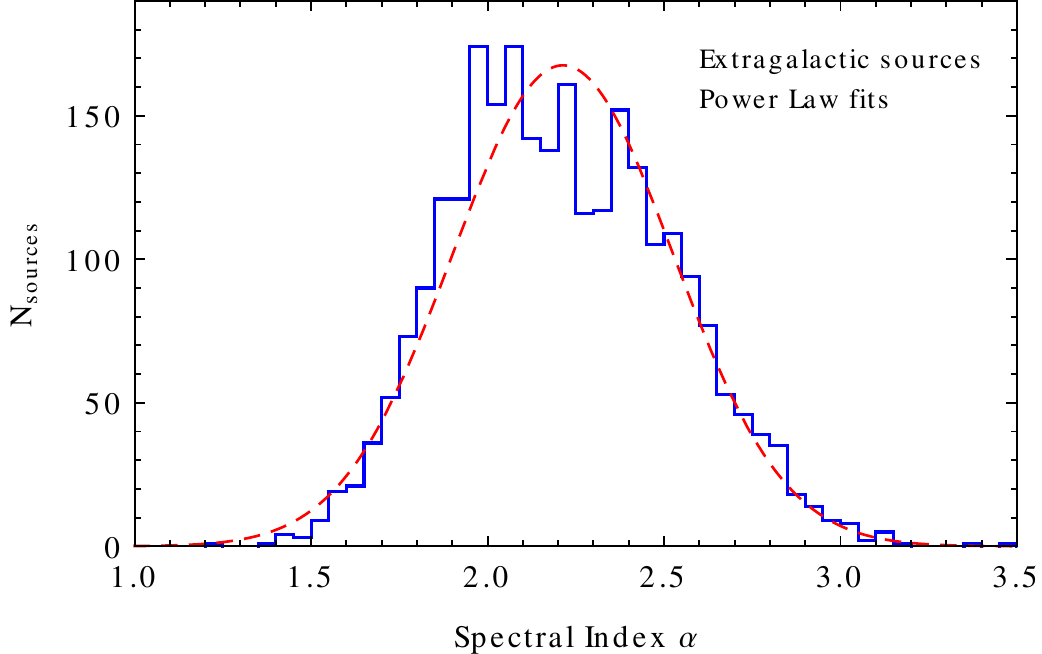}
\end{center}
\caption {\footnotesize
 Distribution of the spectral index $\alpha$ for the
 extragalactic gamma--ray sources in the 4FGL Fermi catalog
 that are in the sky region ($|\sin b | > 0.25$), and 
 have been fitted with a simple power--law form. 
 The thin line is a gaussian with the same average and width of
 the distribution ($\langle \alpha \rangle = 2.21$ and
 $\sigma_\alpha = 0.31$).
 \label{fig:alphapower}}
\end{figure}

Only one of these sources (the blazar 3C 454.3, 
the brightest object in this sky region) is fitted with the ``Cutoff' form.
Most of the selected sources (2629 or 82\% of the total) are
fitted with the simple power--law form.
The spectral index for these objects
has a broad distribution
(shown in Fig.~\ref{fig:alphapower})
that takes values in an interval that extends from 1.2 to 3.5,
with average $\langle \alpha \rangle \simeq 2.21$
and width (r.m.s.) $\sigma_\alpha \simeq 0.31$.

The remaining 593 sources (18\% of the total) are fitted
with the ``log--parabola'' form.
Integrating in the 1--100~GeV energy interval,
these sources account for over 60\% of the total flux.
This is the consequence of the fact that 
sources fitted with the log--parabola
form are on average more luminous than sources fitted
with the simple power--law form.
As an illustration, ranking in total flux the extragalactic sources,
after 3C 454.3 the next 86 objects are
all fitted with the log--parabola form, before
finding the first source fitted with a simple power--law form
(therefore in Fig.~\ref{fig:gamma_resolved} all the best fits lines
are curved).

This result can be (at least in part) attributed to the fact that
it is difficult to measure the curvature of a faint spectrum, and
one interesting hypothesis is that all (or nearly all) extragalactic sources
deviate from the simple power--law form, so that longer observation times
(or more sensitive telescopes) will result in the detection
of a significant curvature for a larger and larger fraction of the sources.

\begin{figure}[bt]
\begin{center}
\includegraphics[width=15.0cm]{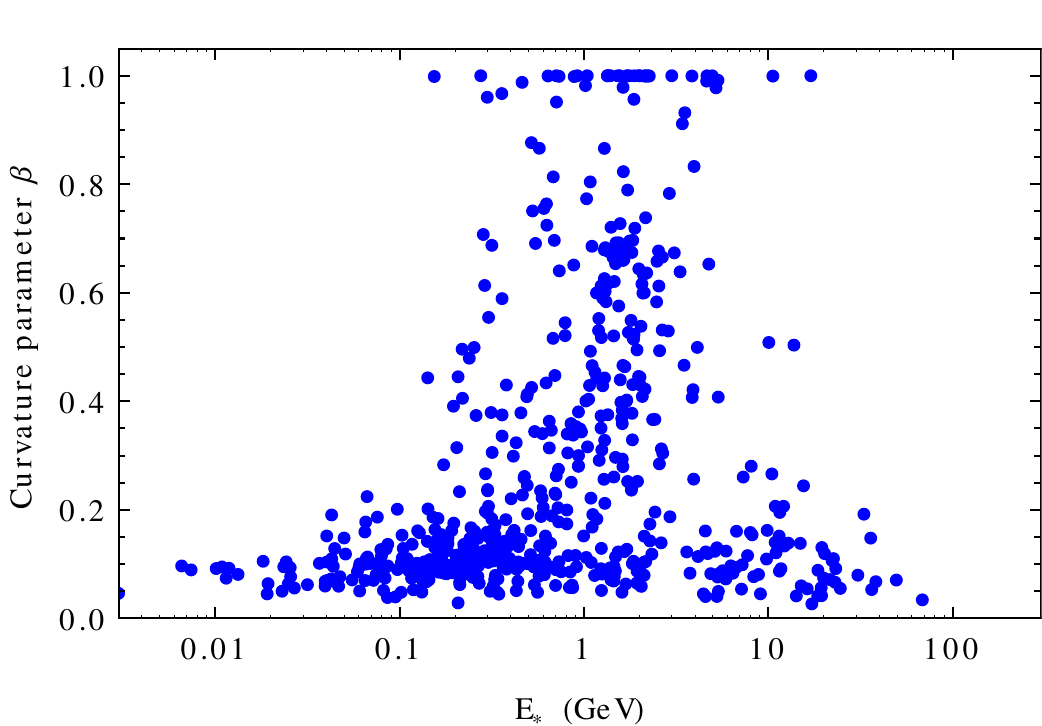}
\end{center}
\caption {\footnotesize
Scatter plot of the shape parameters $E_*$ and $\beta$
 [see Eq.~(\ref{eq:log_parabola_form1})]
for the extragalactic gamma--ray sources of the 4FGL catalog
in the sky region $|\sin b | > 0.25$ 
that have been fitted with the log--parabola form.
The projections of the scatter plot are shown in Figs.~\ref{fig:estar}
and~\ref{fig:beta}.
\label{fig:estar_beta}}
\end{figure}

\begin{figure}[bt]
\begin{center}
\includegraphics[width=12.0cm]{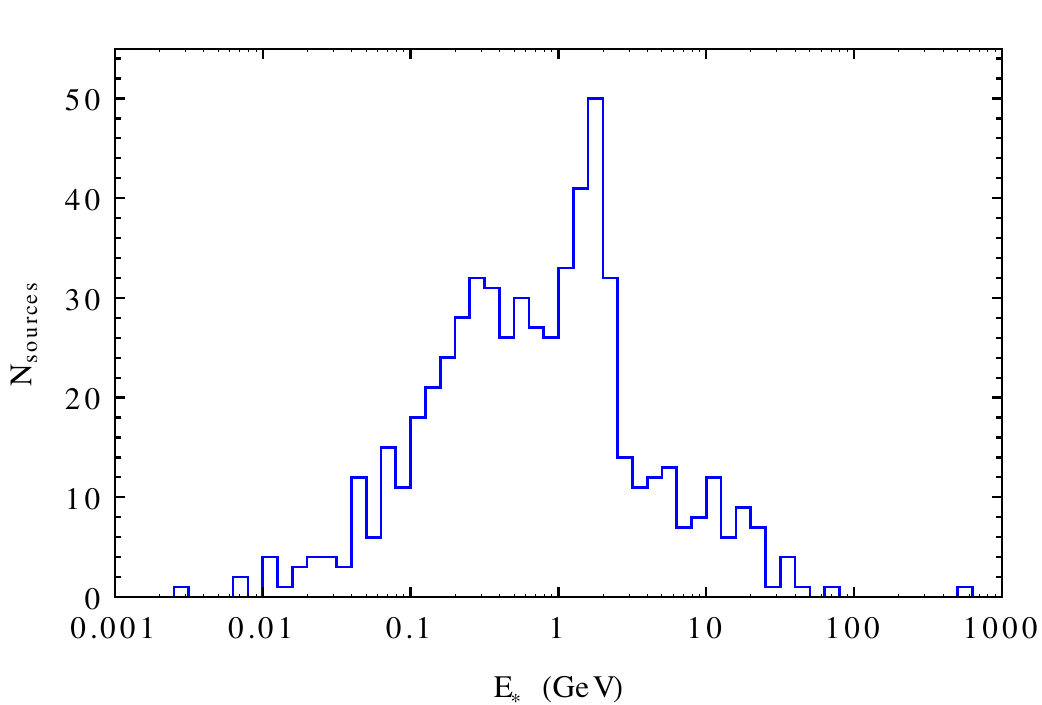}
\end{center}
\caption {\footnotesize
Distribution of the characteristic energy $E_*$
for the extragalactic gamma--ray sources of the 4FGL catalog
in the sky region $|\sin b | > 0.25$ 
that have been fitted with the log--parabola form.
 \label{fig:estar}}
\end{figure}

\begin{figure}[bt]
\begin{center}
\includegraphics[width=12.0cm]{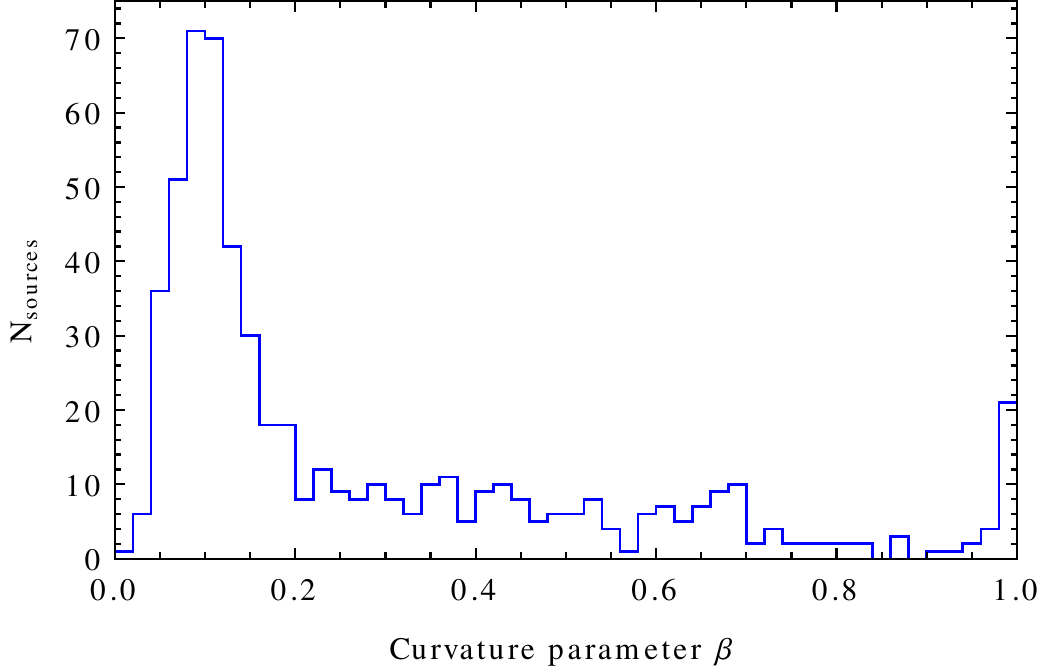}
\end{center}
\caption {\footnotesize
Distribution of the curvature parameter $\beta$
for the extragalactic gamma--ray sources of the 4FGL catalog
in the sky region $|\sin b | > 0.25$ 
that have been fitted with the log--parabola expression.
In the fits the parameter $\beta$ can only take values
in the interval $0 < \beta
\le 1$.
\label{fig:beta}}
\end{figure}

A scatter plot of the shape parameters $\{E_* ,\beta\}$
for the 593 extragalactic sources in sky region $|\sin b| < 0.25$
that have been fitted with the log--parabola form
is shown in Fig.~\ref{fig:estar_beta}.
The projection that gives the distribution of the characteristic
energy $E_*$ is shown in Fig.~\ref{fig:estar},
and one can see that $E_*$ takes values in a broad interval
that extends from 10~MeV to 100~GeV. The projection for
the curvature parameter $\beta$ is shown in Fig.~\ref{fig:beta}.
The distribution has a peak at $\beta \simeq 0.1$, but extends
to values of order unity. Several sources are fitted to the 
maximum value ($\beta = 1$) allowed in the Fermi--LAT fits.

\begin{figure}[bt]
\begin{center}
\includegraphics[width=13.0cm]{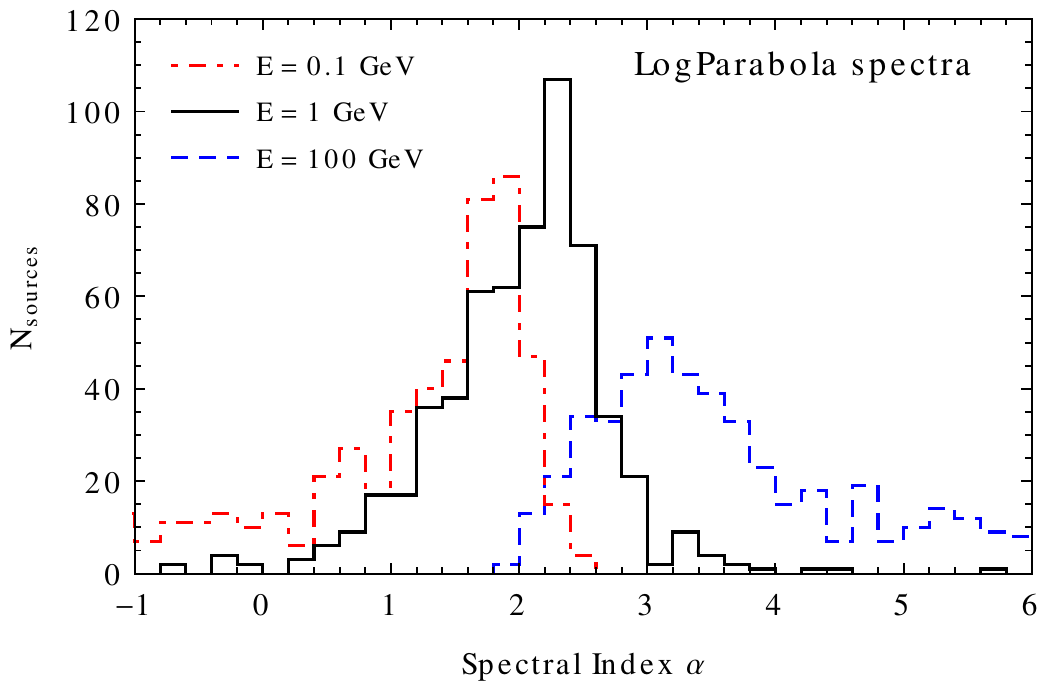}
\end{center}
\vspace{0.2cm}
\begin{center}
\includegraphics[width=13.0cm]{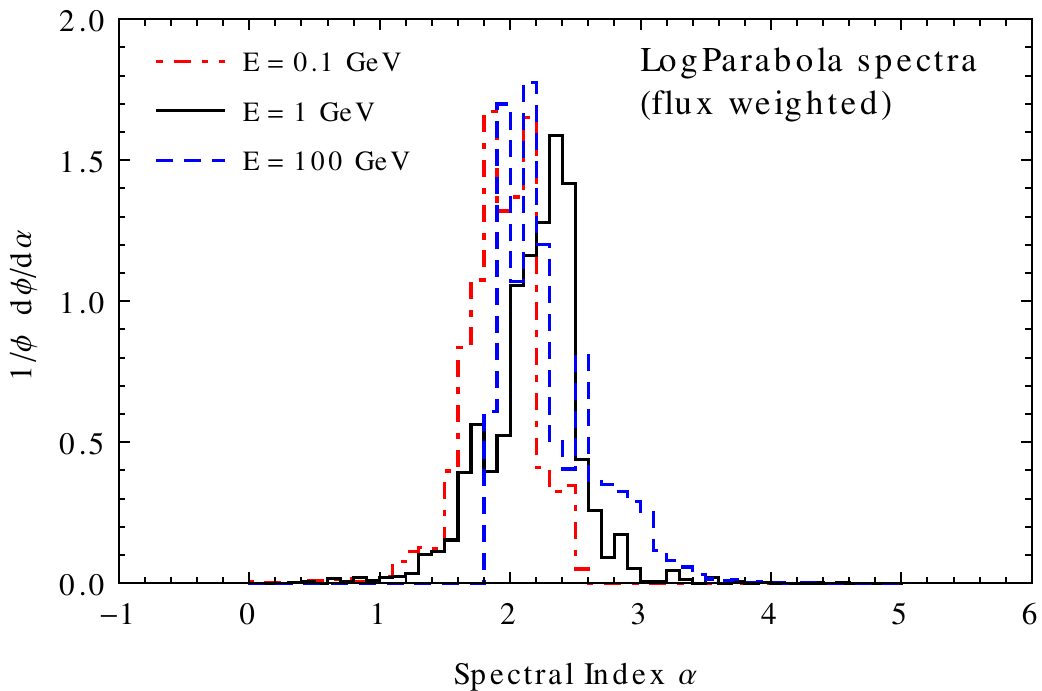}
\end{center}
\caption {\footnotesize
Top panel: Spectral index distribution for
the gamma--ray extragalactic sources of the 4FGL catalog
in sky region $|\sin b| \ge 0.25$ that have
been fitted with the log--parabola form.
The three histograms correspond to energy $E = 0.1$, 1 and 100~GeV.
Bottom panel: as in the top panel, but the contribution
of each source is weighted with the value of the flux
(at the energy considered).
 \label{fig:alpha1}}
\end{figure}

The spectral index distributions of the extragalactic sources
that have been fitted with the log--parabola form
is shown in Fig.~\ref{fig:alpha1}.
The top panel of the figure shows the distribution
of $\alpha(E)$ for three values of the energy
$E= 0.1$, 1 and 100~GeV.
Inspecting this figure one can notice some interesting facts. \\
(i) The shape of the spectral index distribution
is energy dependent. This is because for
fluxes described by the log--parabola expression, 
the spectral index (for $\beta > 0$)
grows with energy. Accordingly, the average $\langle \alpha (E) \rangle$
changes rapidly with energy, taking values of 0.61, 1.96 and 4.6
for $E = 0.1$, 1 and 100~GeV. \\
(ii) The distributions are very broad with r.m.s. values
$\sigma_\alpha (E)$ = 1.74, 0.79 and 2.38 for the same values of the energy.

The bottom panel of Fig.~\ref{fig:alpha1} shows the
distributions of $\alpha(E)$ calculated weighting the contribution
of each source with its (energy dependent) flux.
These flux weighted distributions
have a shape that depends weakly on energy and
is also much more narrow.
For the same three values of the energy used before
($E = 0.1$, 1 and 100~GeV) the average spectral index is
$\langle \alpha (E) \rangle = 1.91$, 2.18 and 2.30,
and the r.m.s. width is $\sigma_\alpha (E) = 0.30$, 0.36 and 0.37.

\begin{figure}[bt]
\begin{center}
 \includegraphics[width=15.0cm]{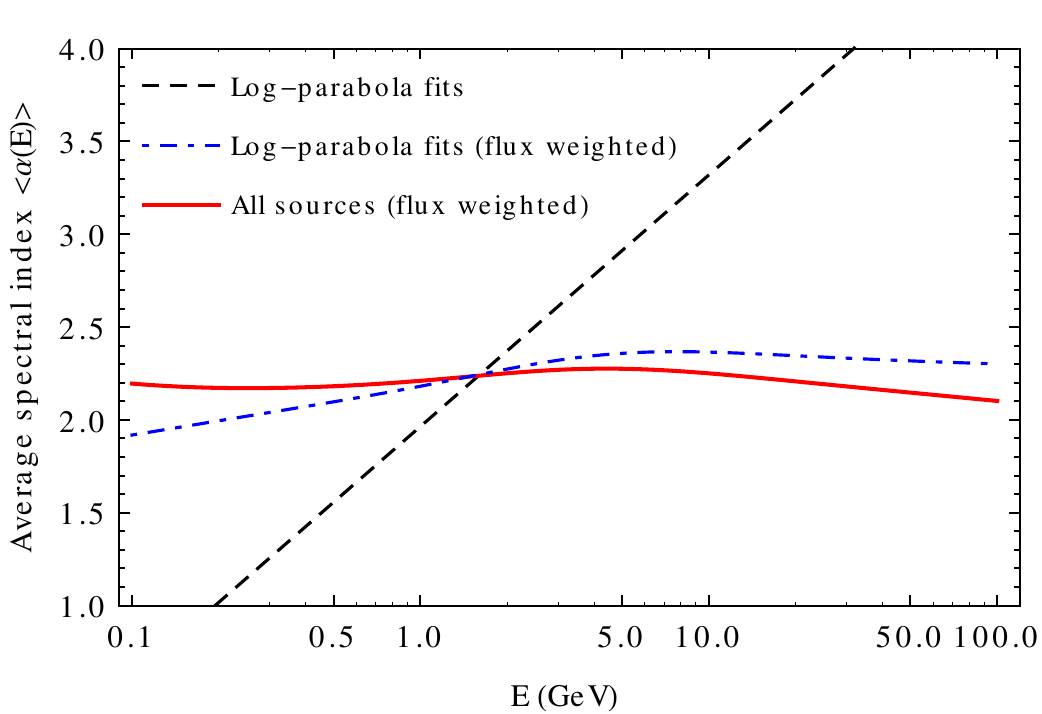}
\end{center}
\caption {\footnotesize
 The (black) dashed line shows, as a function of the energy
 $E$ the average value $\langle \alpha (E) \rangle$
 of the spectral index for the 593 extragalactic sources of the 4FGL catalog
 in the sky region $|\sin b| > 0.25$ that have been fitted with the
 log--parabola expression.
 The (blue) dot--dashed line shows the average spectral index
 for the same set of sources (with log--parabola fits) calculated
 calculated weighting the contribution of each source with the
 (energy dependent) value of the flux.
 The (red) solid line shows the flux weighted average spectral index
 for all extragalactic sources, including also sources fitted
 with a simple power--law form.
 \label{fig:alphamed}}
\end{figure}

The very different energy dependence of the average spectral
index calculated with and without weighting each source
by its flux is also evident in Fig.~\ref{fig:alphamed}.
In the figure the dashed line
shows (as a function of energy) the average
$\langle \alpha (E) \rangle_{\rm sources}$ of the index for all
extragalactic sources fitted with the log--parabola expression.
This quantity grows linearly with $\ln E$ with a coefficient
$2 \, \langle \beta \rangle \approx 0.59$.
The solid line shows the flux weighted average
$\langle \alpha (E) \rangle_{\rm flux}$
of all extragalactic sources (including those with a power law fit).
In this case, in the interval 0.1--100~GeV,
the average takes values in small interval
(between 2.21 and 2.30), and taking into account statistical uncertainties
is consistent with being constant.
This is equivalent to the result that the average of the fluxes
of the extragalactic sources can be described as a simple power--law
in the energy interval considered.

\subsection{Sum of the spectra of the extragalactic point sources}
The difference between the EGB (the total extragalactic flux)
and the IGRB (the isotropic component of the flux) is equal by definition
to the angle averaged contribution of all resolved point sources.
This relation can be written explicitely as:
\begin{equation}
\phi_{\rm resolved} (E) = \phi_{\rm EGB} (E) - \phi_{\rm IGRB} (E) = \frac{1}{\Delta \Omega} ~\sum_j \phi_j(E)
\label{eq:resolved}
\end{equation}
where the summation is over all extragalactic sources in the solid angle
$\Delta \Omega$. As already discussed,
in the energy range 0.1--100~GeV both the EGB and the IGRB
are well fitted by power--law with approximately equal spectral indices
(of order $\alpha \simeq 2.30$). This implies that the average spectrum
of the resolved extragalactic point sources, in reasonably good approximation, 
can also be described by a simple power--law.

\begin{figure}[bt]
\begin{center}
\includegraphics[width=15.0cm]{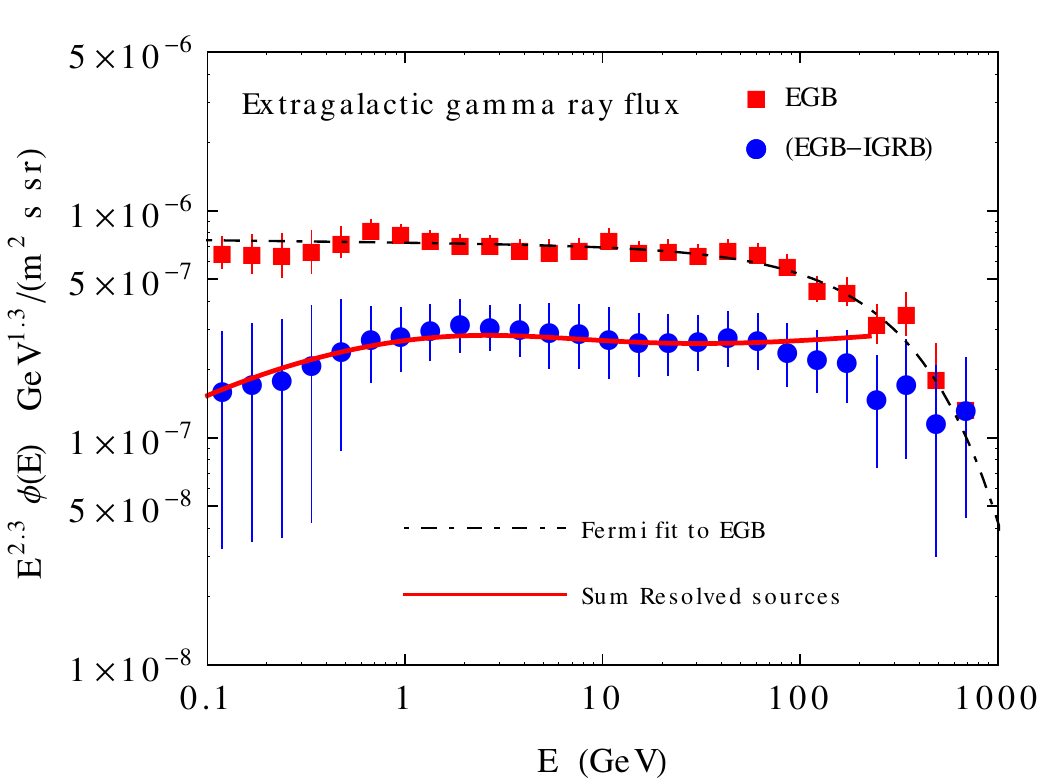}
\end{center}
\caption {\footnotesize
The square points are the Extra Galactic Background (EGB)
measured by the Fermi telescope, plotted as a function of the gamma--ray energy.
The circles give the difference between the EGB and the
Isotropic Gamma Ray Background (IGRB) (that is the component of the EGB
due to resolved sources).
The thick solid line is an estimate of the flux per unit solid angle
of the extragalactic sources measured by Fermi, calculated
summing the fits to all the (3223) sources of the
4FGL catalog in the sky region $|\sin b| > 0.25$ 
divided by the solid angle ($3 \pi$) of the region. 
The dot--dashed line is the fit to the EGB calculated by Fermi.
\label{fig:gamma_extragal}}
\end{figure}

We have checked the validity of Eq.~(\ref{eq:resolved})
obtaining the left--hand size of the equation from the measurements of the EGB and IGRB 
in the Fermi--LAT publication \cite{Ackermann:2014usa}
and calculating the right--hand side summing the fits to
the 3543 extragalactic sources in the sky region $|\sin b| > 0.25$,
and dividing by the appropriate solid angle ($\Delta \Omega = 3 \, \pi$).
The result of this exercise is shown in Fig.~\ref{fig:gamma_extragal},
and shows good agreement in the energy range 0.1--100~GeV.
At higher energy the sum of the fits to the point sources
become larger than the measurement of the resolved extragalactic flux,
but this is expected because the form of the fits does not take 
into account the effects of gamma--ray absorption
that become significant in this range.

The result that the average of the fluxes of all extragalactic point sources
is well described by a simple power--law was in fact already evident
in Fig.~\ref{fig:alphamed} that shows the energy dependence
of $\langle \alpha(E) \rangle_{\rm flux}$, the (flux weighted)
average spectral index of the extragalactic sources.
This quantity is in fact identical to the spectral index
of the sum (or average) of all components, and therefore
the result that $\langle \alpha(E) \rangle_{\rm flux}$ is approximately
constant is equivalent to the statement that the spectrum of the
resolved extragalactic flux is a simple power law.

The result shown in Fig.~\ref{fig:gamma_extragal} is of course
only a consistency check, a demonstration that the fits performed
by the Fermi--LAT Collaboration for the spectra of the detected point sources
are reasonably accurate.
The interesting point is to understand the origin
of this result, that the sum of contributions that have
a large variety of spectral shapes
combine to form a simple power--law spectrum.
This can considered as a simple ``just so'' fact,
but in the following we will take the point of view that
it is something that requires a critical discussion
and the development of an understanding.

It is in fact not obvious how a 
a featureless, scale free power--law spectrum can emerge from
the combination of components that have different shapes.
Also in the case where the components
have power--law form, with a distribution of different slopes,
the average is a convex, gradually hardening spectrum.
The next section will show under which conditions an ensemble
of curved, softening spectra (such as those described by the
log--parabola form) can combine to form an average of power--law form.

\section{Combination of components}
\label{sec:combined-spectra}

\subsection{Power--law components}
\label{sec:combined-power-laws}
It is straightforward to see that the sum of components that
have power--law form but different slopes results in a spectrum
that has a convex, gradually hardening form.

Let us consider a flux that is formed by the sum of many components:
\begin{equation}
 \phi(E) = \sum_j \phi_j (E) = \sum_j K_j ~\left ( \frac{E}{E_0} \right )^{-\alpha_j}
\label{eq:flux-comb0}
\end{equation}
(with $E_0$ an arbitrary reference energy).
The spectral index $\overline{\alpha} (E)$ of the total flux
is simply the average of the spectral indices of the components:
\begin{equation}
\overline{\alpha}(E) = \langle \alpha (E) \rangle = \frac{1}{\phi(E)} ~\sum_j \phi_j (E) \, \alpha_j
 = \frac{1}{\phi(E)} ~\sum_j K_j ~\left( \frac{E}{E_0} \right )^{-\alpha_j} \, \alpha_j ~.
\label{eq:alpha_pow}
\end{equation}
With growing $E$ the contributions of the hard components (with small $\alpha_j$)
increase in importance, because they are weighted with a larger
factor ($\propto E^{-\alpha_j}$), and the spectral index
decreases monotonically with $E$.

Eq.~(\ref{eq:flux-comb0}) can also be rewritten in the form:
\begin{equation}
 \phi(E) = \int d\alpha ~K(\alpha, E_0) ~ \left ( \frac{E}{E_0} \right )^{-\alpha} 
\label{eq:flux-comb}
\end{equation}
where $K(\alpha, E_0)$ gives the contribution to the total flux
at the energy $E_0$ of all sources that have spectral index $\alpha$.
An instructive case is when $K(\alpha, E_0)$ is a gaussian of width $\sigma_\alpha$
around the average value $\alpha_0$:
\begin{equation}
 K (\alpha, E_0) = \frac{\phi(E_0)}{\sqrt{2 \, \pi} \, \sigma_\alpha}
 ~\exp \left [-
 \frac{(\alpha - \alpha_0)^2}{2 \, \sigma_\alpha^2} \right ] ~.
\label{eq:alpha_dist1}
\end{equation}
In this case the integral in Eq.~(\ref{eq:flux-comb})
can be performed analytically
with the result:
\begin{equation}
\phi(E) = \phi(E_0) ~\left ( \frac{E}{E_0} \right )^{-\alpha_0 + \frac{1}{2} \sigma_\alpha^2 ~\ln (E/E_0)} ~.
\label{eq:power-comb}
\end{equation}
The spectral index $\overline{\alpha}(E)$ of the total flux is then:
\begin{equation}
\overline{\alpha} (E) = \alpha_0 - \sigma_\alpha^2 ~\ln \frac{E}{E_0} ~.
\label{eq:alpha-gaussian}
\end{equation}

Comparing Eqs.~(\ref{eq:power-comb}) and~(\ref{eq:log_parabola_form}) 
one can see that an ensemble of components of power law form
with a (flux weighted) distribution of spectral index that
is a gaussian of width $\sigma_\alpha$, combine to form
an average that is of log--parabola spectrum with a (negative)
curvature parameter $\beta = -\sigma_\alpha^2/2$.

In our discussion we have made the assumption that
the distribution of spectral index of the components
is a gaussian at the reference energy $E_0$. This assumption
however implies that the distribution of spectral index is a gaussian
with a constant (energy independent) width $\sigma_\alpha$
for {\em all} values of the energy $E$.
The average $\langle \alpha (E) \rangle = \overline{\alpha}(E)$
does however vary with $E$ following Eq.~(\ref{eq:alpha-gaussian}).
This corresponds to the fact that at different values of the energy,
different components are dominant in forming the total flux,
with harder components becoming important at higher energy.
At the energy $E$, only the subset of sources
with spectral index in an interval of width $\sigma_\alpha$
around the central value
$\langle \alpha (E)\rangle = \alpha_0 - \sigma_\alpha^2 \, \ln(E/E_0)$
give significant contributions to the total flux.

\subsection{Log--parabola components}
\label{sec:combined-logparabola} 

If a spectrum is formed by the combinations of components that
have log--parabola form, the spectral shape is determined
by the combination of two effects that act in opposite directions.
When the energy increases:
(i) components with harder and harder spectral shape
become dominant, but
(ii) the spectra of all components (assuming $\beta > 0$) soften gradually.
It is therefore possible that the two effects cancel resulting
in an average flux that is a simple power--law.

The spectral index of a the sum of an ensemble of
components of log--parabola form around the
(arbitrary) energy $E_0$ can be estimated as:
\begin{equation}
 \overline{\alpha} (E)
 = \langle \alpha(E) \rangle
 \simeq \langle \alpha(E_0) \rangle 
 + \left (2 \, \langle \beta (E_0) \rangle - \sigma^2_{\alpha} (E_0) \right )
 ~\ln \left (
 \frac{E}{E_0} \right )~.
\label{eq:cancellation0}
\end{equation}
In this equation
$\langle \alpha (E_0) \rangle$ 
is the (flux weighted) average spectral index 
at the energy $E_0$, and the curvature of the spectrum
(that is the derivative $d\overline{\alpha}(E)/d\ln E$)
is determined by the sum of two terms of opposite sign
that describe the two effects discussed in the previous paragraph.
The first one: 2 $\langle \beta(E_0)\rangle$
is associated to the fact that the spectra of the individual
components are softening [see Eq.~(\ref{eq:alpha_logparabola1})],
and is simply the average of the curvature parameters of the components.
The second term 
is associated to the fact that
at higher energy the total flux is formed by harder components
[see Eq.~(\ref{eq:alpha_pow})], and $\sigma_{\alpha} (E_0)$
is the width of the spectral index distribution at $E_0$.

Inspecting Eq.~(\ref{eq:cancellation0}) one can see that the curvature of the
spectrum of the average flux at the energy $E$ can
vanish if the condition
\begin{equation}
2 \, \langle \beta(E) \rangle = \sigma_\alpha^2 (E)
\label{eq:cancellation}
\end{equation}
is satisfied.
If Eq.~(\ref{eq:cancellation}) is valid in a finite energy interval,
then in that range the spectrum is described by a simple power--law
of constant spectral index.

The spectrum of the resolved extragalactic gamma--ray flux
obtains its power law form because of this ``cancellation effect''.
At higher energy the flux is generated by harder components,
however the spectra of the individual components are also
gradually softening, and the spectral index of the total flux
remains approximately constant.

\section{Statistical properties of the ensemble of the gamma--ray sources}
\label{sec:statistical}

The fact that the extragalactic gamma--ray sources
emit spectra that have a variety of different shapes,
is of course of great importance
to develop an understanding of the mechanisms that generate the spectra.
It is however not immediately clear what is the significance of the fact that
the average flux formed by the ensemble of all sources can be well
described by a simple power law in a broad energy range.
This result can be seen as an 
``accident'' without any deep physical meaning.
In the following however we will start from the assumption that this
result is perhaps revealing some interesting property for the
{\em ensemble} of the extragalactic sources.

We will discuss this problem in terms an ensemble of sources
that emit spectra $q_j(E)$ (with $j$ an index that runs over all the sources)
of different shape and normalization.
The combined emission, obtained summing all sources will be indicated by $Q(E)$.

A result of general validity is the following.
The combined emission $Q(E)$ has a simple power--law form if
these two conditions are satisfied:
\begin{itemize}
\item [(A)] The spectral shape of each component
is determined by a parameter $E_*$ (with the dimension of energy),
and is a function of the ratio $E/E_*$, so that 
the emission from the $j$--th source has the form:
\begin{equation}
 q_j (E) = \frac{q_{0,j}}{E_{*,j}} ~F\left ( \frac{E}{E_{*,j}} \right )
\label{eq:qj}
\end{equation}
where $q_{0,j}$ is a normalization factor,
$E_{*,j}$ is the (source dependent) characteristic energy,
and $F(x)$ is a function of arbitrary shape.
Without loss of generality one can impose the condition
that $F$ is normalized to unity,
so that the energy integrated emission from the $j$--th source
is $q_{{\rm tot},j} = q_{j,0}$.
\item[(B)] The (energy integrated) emission 
 of all sources characterized with characteristic energy
 $E_*$ has the power law dependence:
\begin{equation}
\frac{dQ_{\rm tot}}{dE_*} (E_*) = Q_0 ~\left (\frac{E_*}{E_0} \right )^{-p} 
\label{eq:qtot-estar}
\end{equation}
(with $E_0$ an arbitrary reference energy).
For $p > 0$ Eq.~(\ref{eq:qtot-estar})
states that the emission of the sources
decreases when their spectral hardness increases.
\end{itemize}

A demonstration of the theorem stated above is straightforward.
The total emission $Q(E)$ can be written as an integral
over the parameter $E_*$ as:
\begin{equation}
 Q(E) = \sum_j q_j (E) = \int_0^\infty dE_* 
 ~\frac{dQ_{tot}}{dE_*}
 ~\frac{1}{E_*}
 ~F\left ( \frac{E}{E_{*}} \right )
\label{eq:q0}
\end{equation}
where the quantity $dQ_{\rm tot}/dE_*$
\begin{equation}
 \frac{dQ_{\rm tot}}{dE_*} = \sum_j q_{0,j}~\delta[E_* - E_{*,j}]
\label{eq:dqstar}
\end{equation}
describes the contribution to the emission
of all sources with critical energy $E_*$.
If $dQ_{\rm tot}/dE_*$ has the power--law form of Eq.~(\ref{eq:qtot-estar})
the integration over $E_*$ in Eq.~(\ref{eq:q0})
can be performed analytically, and the combined emission
takes the form:
\begin{equation}
Q(E) = Q_0 \; k_p ~\left (\frac{E}{E_0} \right )^{-p} 
\label{eq:q1}
\end{equation}
where $k_p$ is an adimensional constant that depends
on the exponent $p$ and on the shape of the function $F(x)$:
\begin{equation}
k_p = \int_0^{\infty} \; dx ~x^{p-1} \; F(x) ~.
\label{eq:kk}
\end{equation}
This completes the demonstration of our general theorem.
The important point of Eq.~(\ref{eq:q1}) is that the
spectral index $p$ of the combined emission
is associated to the statistical property of
the ensemble of the sources,
and describes how the luminosity of the sources
decreases with the hardness of their spectra.

\subsection{Log--parabola spectra}
The general discussion that we have developed above can be applied
to spectra of log--parabola (or log--normal) spectra.
The only complication is that the log--parabola
spectra are defined not by a single parameter, but by two,
that can be chosen as the characteristic energy $E_*$
and the curvature parameter $\beta$
(see discussion in Sec.~\ref{sec:point-sources}).

If $\beta$ has a single value the results discussed above
are immediately applicable. In this case the (normalized) function $F(x)$
[see Eq.~(\ref{eq:log_parabola_form1})] has the form:
\begin{equation}
 F(x) = \sqrt{\frac{\pi}{\beta}} ~e^{-1/(4 \, \beta)} ~x^{-(2+\beta \, \ln x)}
\label{eq:flog}
\end{equation}
and the constant $k_p$ in (\ref{eq:kk}) becomes:
\begin{equation}
k_p = e^{(p^2-4 p + 3)/(4 \, \beta)}
\label{eq:kk-log}
\end{equation}

The quantity $dQ(E)/dE_*$ that describes the contribution
of sources characterized by the parameter $E_*$
to the total emission at the energy $E$
can be calculated using Eqs.~(\ref{eq:q0}) and~(\ref{eq:flog})
with the result:
\begin{equation}
 \frac{1}{Q(E)} ~\frac{dQ(E)}{d\ln E_*} =
 \frac{1}{\sqrt{2 \, \pi} \; \sigma_{\ln E_*}} ~
 \exp \left [
 - \frac{ (\ln E_* - \langle \ln E_* (E) \rangle)^2} {2 \, \sigma_{\ln E_*}^2} 
 \right ] ~.
\label{eq:q-logestar}
\end{equation}
This distribution is a gaussian in $\ln E_*$ with width 
\begin{equation}
\sigma_{\ln E_*} = 1/\sqrt{2 \, \beta}
\end{equation}
and average 
\begin{equation}
\langle \ln E_* (E) \rangle = \ln E - \left (\frac{p-2}{2 \, \beta} \right ) ~.
\end{equation}
In other words, the emission at the energy $E$ is generated by
sources that have the $E_*$ parameter in a relatively small range of
values centered at a value $E_* \simeq E \, e^{-(p-2)/(2 \, \beta)}$
that grows linearly with $E$, so that increasing the energy, the emission is
dominated by harder and harder sources
(with larger and larger characteristic energy $E_*$).

The combined emission at energy $E$ has the spectral index $p$,
but the slopes of the contributions of the individual sources have a range
of values. It is easy to compute the shape of this distribution
using Eq.~(\ref{eq:q-logestar}) and the relation between the slope
$\alpha(E)$ and the critical energy $E_*$. The result is:
\begin{equation}
 \frac{1}{Q(E)} ~\frac{dQ(E)}{d\alpha} =
 \frac{1}{\sqrt{2 \, \pi} \; \sigma_{\alpha}} ~
 \exp \left [
 - \frac{ (\alpha - p)^2} {2 \, \sigma_{\alpha}^2} 
 \right ] ~.
\label{eq:q-alpha}
\end{equation}
This expression is again a gaussian, however in this case
both the width and the average are energy independent.
The average spectral index is 
\begin{equation}
\langle \alpha \rangle = p
\end{equation}
(as it must because we have already demonstrated that the
combined emission is a power law of constant slope), and the
constant width is again:
\begin{equation}
\sigma_\alpha = 1/\sqrt{2 \, \beta} ~.
\end{equation}

The results of Eqs.~(\ref{eq:q-logestar}) and~(\ref{eq:q-alpha})
have been obtained for a unique value of the curvature parameter $\beta$,
it is however straightforward to generalize to the case
where the sources have an arbitrary $\beta$ distribution
with a shape that is independent from the value of
the characteristic energy $E_*$,
as (in first approximation) is the case for the extragalactic gamma--ray
sources (see Fig.~\ref{fig:estar_beta}).
In this more general case the distributions of $\ln E_*$ and
of $\alpha(E)$ are the superposition of gaussians with
$\beta$ dependent width and average.
The distributions of $\alpha(E)$ of the extragalactic sources
shown in the bottom panel of Fig.~\ref{fig:alpha1}
have an energy independent shape
and are consistent with the results of Eq.~(\ref{eq:q-alpha}).

\section{The origin of power--law spectra}
\label{sec:power-laws}
Power--laws appear widely in
physics, biology, economics, social sciences and many other fields.
For instance they describe the distributions of the size of earthquakes,
moon craters, towns and cities, forest fires and
many other data sets \cite{newman_powerlaws}.
A list of real--world data sets from a range of different
disciplines that can be reasonably well described by
power--law distributions is for example presented in \cite{clauset_powerlaws},
and it is remarkable that the spectral indices
that describe these data sets have values between 1.7 and 3.1,
and in several cases the best fit to the spectral index
is of order 2.3--2.4.
The origin of these power--law distributions 
has been a topic of debate for over a century.

A possible explanation for a number of these power--laws is 
the intriguing concept of self--organized criticality, originally proposed
by Bak {\it et al.} \cite{Bak:1987xua,Bak:1988zz}
to describe dynamical systems (such as the paradigmatic sandpile model)
that evolve naturally toward a critical state that has no intrinsic time or length scale.
Well known ``toy models'' examples of this idea
have been developed with numerical simulations of cellular automata,
that describe approximations of sand piles \cite{Bak:1988zz}
or forest fires \cite{Bak:1990,Drossel:1992}.
The concept of self--organized criticality has been applied to a wide range
of fields from biophysics (evolution and extinctions, spread of diseases)
to social sciences (urban growth, traffic, internet),
and also in astrophysics \cite{Aschwanden:2014dna}.

It is not clear if the concept of self--organized criticality is
also relevant to understand the origin of the approximate power--law shape of
the cosmic ray spectra, a problem that has been of central importance
in high energy astrophysics for many decades.

Current models for Galactic cosmic rays
(see for example the textbook of
Gaisser, Engel and Resconi \cite{Gaisser:2016})
explain the power--law form of the spectrum with the existence
of a ``universal'' acceleration mechanism
that generates power--law spectra with a unique spectral index $\alpha$
(below a maximum energy that can be source dependent).
The source spectra are then distorted by propagation effects,
that for ultrarelativistic protons and nuclei have
a power--law energy dependence
characterized by the slope $\delta$, so that 
the CR fluxes observable at the Earth have power--law spectra
with a slope $\gamma$ determined by both 
(acceleration and propagation) mechanisms: $\gamma \simeq \alpha + \delta$.

The ``universal'' CR acceleration mechanism is commonly identified 
as ``first order Fermi acceleration'' \cite{Blandford:1987pw}
(based on a modification of ideas originally proposed by Fermi \cite{Fermi:1949ee}),
where particle are accelerated while propagating in a magnetized
plasma in the presence of strong shock waves,
such as those generated by supernova explosions.
Simplified treatments of this mechanism
generate spectra with index $\alpha = 2 + 4/M^2$ where
$M$ is the Mach number of the shock wave in the upstream region.
For strong shocks ($M \gg 1$) this corresponds to an approximately universal
slope $\alpha \simeq 2 +\varepsilon$ (with $\varepsilon$ positive and small).

The ``standard'' model for the acceleration of Galactic cosmic rays
appears quite distant from the concepts of self--organized criticality.
On the other hand, in this work we have shown that, under certain conditions,
a power--law spectrum can also be formed by the combination of
components that have different shapes.
In particular, the discussion in Sec.~\ref{sec:statistical}
demonstrates that the power--law shape of the extragalactic gamma--ray flux
emerges from the relation between the luminosity of the sources
and the hardness of their spectra [see Eq.~(\ref{eq:qtot-estar})].
This luminosity--hardness relation is of power--law form, and its slope
determines the spectral index of the gamma--ray emission.

If the origin of the power--law form of the spectrum
is based on a mechanism of this type, and is related to the
statistical properties of the ensemble of the sources,
the concepts of self--organized criticality
become much more relevant.

For example, an intriguing possibility is that there is an analogy between
the flares that accelerate particles in blazars and earthquakes.
The distribution of the energy released during earthquakes has been found
to obey the well known Gutenberg--Richter law \cite{gutenberg-richter},
originally proposed on the basis of empirical observation.
The Gutenberg--Richter law is usually formulated
stating that the frequency of
earthquakes with magnitude greater than $m$ is given by the relation
$\log_{10} N = a - b\; m$, where $a$ and $b$ are adimensional constants
with values that depend on the region of the Earth
(with the parameter $b$ close to 1.0).
The Gutenberg--Richter law can be reformulated
stating that the differential frequency distribution for the
release of energy $\mathcal{E}$ in an earthquake 
has the power--law form: $dN/d\mathcal{E} \propto \mathcal{E}^{-(b+1)}$.
The mechanisms that generate this frequency--magnitude relation
are not yet fully clarified, but have been interpreted as the
the consequence of the fact that the Earth crust is in a state of
self--organized criticality \cite{Bak:2002zz}. 

One can speculate that the blazar emission (that dominates
the extragalactic gamma--ray flux) is generated by
``flares'' (presumably associated to the accretion
flow on the central black hole) that are less frequent
(or less energetic) when they form harder spectra.
The spectral index of the average extragalactic
flux is then related to the exponent that describes
the frequency of events with high and low characteristic energy.
In this framework, the slope of the average extragalactic flux
is analogous to the parameter $(b+1)$ of the Gutenberg--Richter law.

A better and closer analogy for the origin of the extragalactic gamma--rays
is the acceleration of particles in solar flares.
The spectra of relativistic particle generated in
solar flares have a large variety of spectral shapes, 
however the time averaged spectra 
measured in the energy range from 10~KeV to 100~MeV
\cite{mewaldt_2001,mewaldt_2007} have a smooth shape
that can be well approximated as a simple power--law.
This result can be explained with the same argument
outlined above, and assuming that the generation
of the flares is a critical phenomenon.
In fact already in 1991 Lu and Hamilton \cite{Lu-Hamilton} have argued that 
solar flares can be seen as analogous to the avalanches of sand
in the models published by Bak and colleagues.
This naturally explains why the flares have a very broad size distributions 
that is well described by a single power--law
that spans five orders of magnitude.
In this interpretation the classification of flares
into nanoflares, microflares,
giant flares and so on, is arbitrary, because they
are all generated by the same fundamental mechanism.
In order to form a featureless time integrated spectrum, it is also
necessary to assume that the spectral shape of the particles
accelerated in one flare is related to the total energy contained
in one event.

\section{Galactic cosmic rays} 
\label{sec:galactic}

As discussed above, the study of Galactic cosmic rays indicates
that the average source spectrum released in interstellar space
by the Milky Way accelerators, in a broad energy range, has a power--law shape with
a spectral index of order $\alpha_0 \approx 2.2$--2.4.
In the ``standard model'' for the acceleration of Galactic cosmic rays
all CR sources (or at least those that are dominant)
generate spectra that have a unique, ``universal'' shape
(that is obviously identical to the shape
of the space and time averaged spectrum).

The prediction of the existence of this ``universal'' acceleration
spectrum has very important implications for high energy astrophysics,
but has not yet received a clear confirmation from the observations.
The alternative possibility is that the power--law form of the Galactic
CR source spectrum emerges as the average of components
of different spectral shape.

This question can be investigated experimentally,
studying the energy distributions of freshly accelerated particles
inside or near the acelerators.
Information about these CR spectra can be obtained from the
observations of the emission of gamma--rays (and neutrinos) from these
astrophysical sources.

We have already used in this work the Fermi--LAT observations
of the gamma--ray sources summarized in the 4FGL catalog 
\cite{Fermi-LAT:2019yla} to study the extragalactic sources.
The catalog also  give information about Galactic sources, and
in particular about the spectra of young Supernova remnants (SNR),
that are commonly considered as the most attractive class of objects
for the main accelerators of Galactic cosmic rays.

\begin{figure}[bt]
\begin{center}
\includegraphics[width=15.0cm]{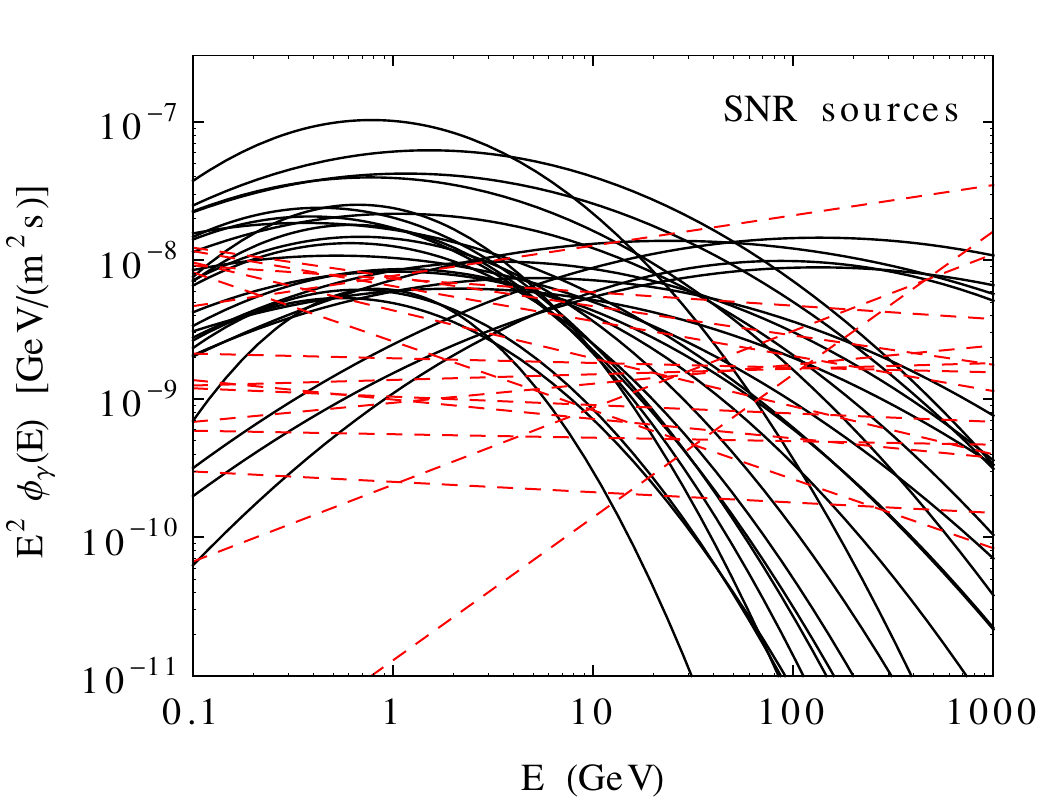}
\end{center}
\caption {\footnotesize
 Spectra of all sources in the 4FGL catalog
 that are classified as supernova remnants (SNR and snr). 
 The solid (black) lines are sources fitted with the log--parabola
 form, the dashed (red) lines are sources fitted
 with a simple power--law form.
 \label{fig:snr_spectra}}
\end{figure}

The 4FGL catalog contains 40 sources that are associated to SNR. 
The best fits for these 40 sources are shown in Fig.~\ref{fig:snr_spectra}. 
Fifteen of these SNR sources have been fitted with
a simple power--law form, and the spectral index of these fits
takes values in a broad interval that goes from 0.96 to 2.49,
with average $\langle \alpha \rangle \simeq 1.98$,
and a width (r.m.s.) $\sigma_\alpha = 0.36$.
The other 25 sources have
curved spectra and have been fitted with the log--parabola
expression of Eq.~(\ref{eq:log_parabola_form}).
Inspecting the figure one can see that sources
fitted with the log--parabola form are typically
brighter than those fitted with the power--law form,
and summed together account for 90.1\% of the flux of all SNR sources.

\begin{figure}[bt]
\begin{center}
\includegraphics[width=15.0cm]{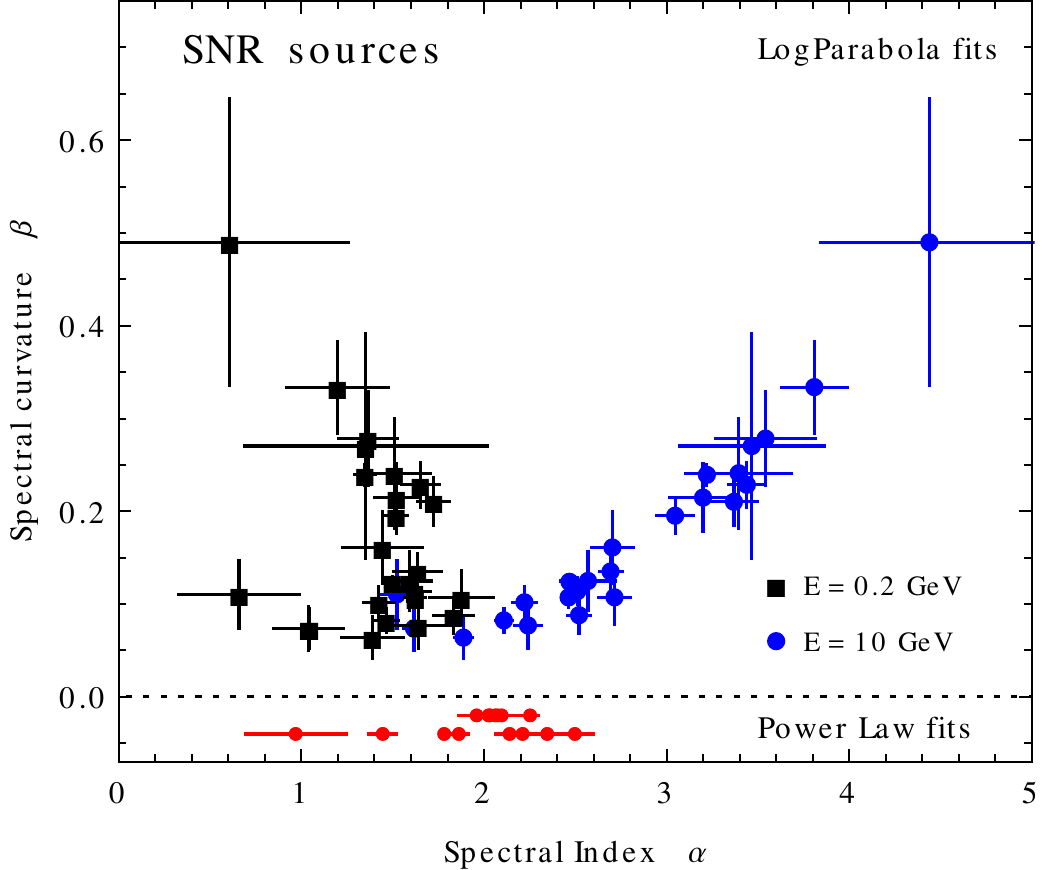}
\end{center}
\caption {\footnotesize
Shape Parameters of the fits to the sources in the 4FGL catalog
 that are classified as supernova remnants (SNR and snr). 
 For the 15 SNR sources that have spectra fitted with a
 simple power--law form the figure shows the (constant) spectral index.
 For the 25 sources fitted with the log--parabola form the figure
 plots the point $\{\alpha_0(E_0),\beta\}$
 (where $\alpha_0(E_0)$ is the spectral index at the energy $E_0$,
 and $\beta$ is the spectral curvature parameter). For each source
 the figure shows the parameters for two values of the energy:
 at $E_0 = 0.2$~GeV and $E = 10$~GeV.
 \label{fig:snr_parameters}}
\end{figure}

Fig.~\ref{fig:snr_parameters} shows the shape parameters of the fits
to the SNR gamma--ray spectra.
For the 15 sources fitted with a power--law form,
the lower part of the figure shows the best fit spectral index
and the 1--$\sigma$ error. For the 25 sources fitted
with the log--parabola form,
the figure shows the spectral index $\alpha(E)$ for two values
of the energy: $E = 200$~MeV and $E = 10$~GeV.
For each of these sources the spectral shape is represented by
two points (with error bars) in the plane $\{\alpha, \beta\}$
(showing the spectral index $\alpha$ at the two energies considered,
while the parameter $\beta$ is energy independent). 

Figures~\ref{fig:snr_spectra} and~\ref{fig:snr_parameters}
show that the gamma--ray emission from SNR has a very broad range of
spectral shapes, and this suggests that the CR populations
contained in the sources have also a broad range of energy distributions.

Can these results be reconciled with the idea that
all SNR generate spectra of universal shape?
Unfortunately the answer to this question is not trivial and requires
a detailed modeling of the sources. There are in fact
some significant difficulties for the 
interpretation of the SNR data.  \\
(i) The observations of each source are effectively only one
``snapshot'', taken at a single time, of an evolving object. \\
(ii) One needs a model for the space distributions of the
populations of relativistic particles and of the target 
(gas and radiation fields) inside the sources.

Given these difficulties, it is perhaps possible that the large
variations in the spectral shape of the emission from SNR
remnants can be attributed to differences in the age and environment
of the supernovae, so that the time integrated spectra of different objects
have equal shape.
In this work we will not attempt a discussion of these problems
and a review of the large body of literature
on the modeling of particle acceleration
in Supernova Remnants. We can however comment that also the
alternative possibility, that different supernovae accelerate
cosmic rays populations with different spectral shapes is consistent with
the observations. The hypothesis that the spectra of particles
accelerated in SNR do not have a unique spectral shape is
not necessarily inconsistent with the idea that 
SNR are the dominant source of the Galactic cosmic rays,
if the contributions of a sufficiently large number of
objects combine to form an average spectrum of power--law form.

Recently, it has been observed that the Galactic CR spectra
do not have exactly a power--law form, but contain
features such as a hardening at a rigidity of order 300~GV, and a
softening at 10~TV \cite{Ahn:2010gv,Adriani:2011cu,Aguilar:2015ooa,An:2019wcw}.
A possible interpretation for these deviations of the spectrum
from a simple power--law form is that they are
the manifestation of the fact that the CR flux is
formed by components that do not have identical
energy distributions \cite{Lipari:2019jmk}.

\section{Outlook}
\label{sec:conclusions}
The observations of Fermi--LAT have shown that most of extragalactic
gamma--ray flux is generated by blazars.
These sources emit spectra that have a broad range of shapes
that in most cases are ``curved'', that is they do not have
a constant spectral index, but soften gradually when the energy increases.
This fact is clearly of great importance to develop an understanding
of the mechanisms that accelerate and confine particles in AGN jets.

The fact that the emissions from blazars do not
have a unique shape implies that the extragalactic
flux at different energies is generated by different objects.
This can be of great importance
for the study of the origin of the astrophysical neutrino flux
recently discovered by IceCube
\cite{Aartsen:2013jdh,Aartsen:2014gkd,Aartsen:2015rwa}.
This neutrino flux emerges as an approximately isotropic
(and therefore extragalactic) component
above the atmospheric foreground at very high energy ($E \gtrsim 100$~TeV).
Blazars, that are the dominant source of extragalactic gamma--rays
in the energy 0.1--10$^3$~GeV, are the most
natural candidate for the class of objects that generates
the neutrino signal, because in all theoretical models
high energy gamma--ray and neutrino emissions are intimately related.
There is strong evidence \cite{IceCube:2018dnn,IceCube:2018cha}
that one blazar (TXS 0506+056) is a high energy neutrino emitter,
however studies of the correlation between the directions of the
high energy neutrinos detected by IceCube and the positions of the
blazars observed by Fermi--LAT have yielded only upper limits
\cite{Aartsen:2016lir} on the maximum contributions of these objects
to the astrophysical neutrino signal.
For these correlation studies it is important to tale into account
the fact that the gamma--rays and the neutrinos are observed
in different energy ranges, and the result discussed in this paper
that different sources have different spectral shapes must
be taken carefully into account to obtain an estimate of the
contribution of the blazars to the neutrino flux.
This problem deserves a detailed discussion
that is postponed to a future paper.

It is remarkable that the average spectrum generated by the ensemble of
all extragalactic sources, in a broad energy interval,
can be well described by a simple, featureless power--law form,
with a spectral index of order 2.30.
This result emerges even if the spectra of the individual sources are
``curved'' and gradually softening, because with increasing energy
objects with harder spectra become dominant. For any $E$
in the range of the Fermi--LAT observations
the extragalactic flux is dominated by sources with 
spectral index of order 2.30. The sources dominant at energy $E$
are less important both at lower energy
(when they have a harder spectra and their
relative contribution is growing) and at higher energy
(when they have softer spectra and their contribution is decreasing).

The ``hardness'' of the spectrum of a source 
can be parametrized by the value $E_*$ of the energy where
the spectral index has the value
$\alpha(E_*) = 2$. The result that the average spectrum of
all extragalactic sources has a simple power--law form is then
equivalent to the statement that the emission $dQ_{\rm tot}/dE_*$ 
of the sources that generate spectra characterized by the parameter $E_*$
has a power--law dependence on $E_*$: $dQ_{\rm tot}/dE_* \propto E_*^{-p}$.
The exponent $p$ is then also the spectral index of the average
emission spectrum.

The emission of gamma--rays from blazars is then analogous
to the production of solar energetic particles, that are generated by
solar flares that have a very broad range of sizes, with small
and frequent flares that generate soft spectra, and large and rare flares
that generate hard spectra. The time averaged spectrum of the flares
is also reasonably well described by a simple power--law.

In this scenario, the power--law shape of the extragalactic gamma--ray
emission emerges because of the statistical properties of the
blazar flares, and the (power--law form) relation between
the flares frequency and energy output, and the hardness of the
spectra of the particles that they accelerate.
The flaring of blazars can then be seen as one example of a critical
phenomenon, analogous for example to the generation of earthquakes in
the crust of the Earth.

It is natural to speculate if some of the results and
considerations developed here for the gamma--ray emission
from blazars can be relevant also for other classes of
high energy sources, in particular for the acceleration
of Galactic cosmic rays.
In this respect it is interesting to note
that also the Galactic gamma--ray sources measured by Fermi--LAT
have a broad range of spectral shapes, and
a large fraction of them is fitted with the
``curved'' log--parabola expression.
The sources with gradually softening spectra are
also bright and account for 73\% of the total flux
of all point sources (excluding Pulsars).
For Supernova remnants, objects fitted with the log--parabola spectrum
account for more than 90\% of the flux. 
Since the curvature of the spectrum of a faint object is
difficult to observe, this suggests that perhaps most
of both Galactic and extragalactic sources have curved spectra.

These results appears in conflict with the simple idea
that astrophysical acceleration mechanisms always
generate power-law spectra,
and suggest to investigate in depth alternative models,
where only the average of many sources can be described by
a spectrum of constant slope.

\vspace{0.25 cm}

\noindent{\bf Acknowledgments.}
I'm grateful to Tom Gaisser for pointing my attention to the
acceleration of particles in solar flares, and to Silvia Vernetto
for many discussions.

\appendix

\section{The origin of the log--parabola spectrum} 
\label{sec:log-parabola-origin}
In this paper we will not attempt an in depth discussion of the mechanisms that 
could generate particles with a log--parabola spectrum,
it can be however interesting to note
that spectra of (approximately) the log--parabola form
can emerge naturally in models
where particle are accelerated with the standard first order Fermi mechanism
introducing some simple modifications.

In the Fermi mechanism the acceleration is a stochastic process where
the charged particles acquire energy gradually in many small steps.
The standard, ``textbook'' treatment of Fermi acceleration
can be summarized as follows.
In each acceleration step (or cycle) a particle increases its energy by
an amount $\Delta E = \xi \, E$ proportional to $E$,
with $\xi$ a small adimensional constant.
The probability that the acceleration process stops
during one cycle has the constant value $P$.
It is straightforward to demonstrate that if particles are injected
in the accelerator with the initial energy $E_0$, the ensemble of particles that
``exit'' from the process has a power--law distribution with spectral index
$\alpha = P/\xi + 1$.

A formal derivation of this well known result can be obtained
approximating the acceleration as a continuous deterministic process,
where the energy $E(t)$ of a particle varies in time as:
\begin{equation}
\frac{dE}{dt} = \frac{\xi \; E}{T_{\rm cycle} (E)} ~,
\label{eq:tau01}
\end{equation}
and describing the probability of stopping the acceleration as a 
continuous loss of particles:
\begin{equation}
\frac{dN}{dt} = -\frac{P \; N}{T_{\rm cycle} (E)} ~
\label{eq:tau02}
\end{equation}
(with $N(t)$ the number of particles inside the accelerator at time $t$).
In these equations $T_{\rm cycle}(E)$ is the time required to complete
an acceleration cycle, that can have an arbitrary energy dependence.
It is convenient to introduce the adimensional variable $\tau$ that counts
the number of acceleration cycles, and is defined by $d\tau = dt/T_{\rm cycle} (E)$.
Equations~(\ref{eq:tau01}) and~(\ref{eq:tau02}) can then be rewritten as:
\begin{equation}
\frac{dE}{d\tau} = \xi \; E
\label{eq:tau1}
\end{equation}
\begin{equation}
\frac{dN}{d\tau} = -P \; N 
\label{eq:tau2}
\end{equation}
The integration of these equations is trivial. 
Particles that enter
the accelerator at $\tau =0$ with energy $E_0$, increase
their energy as $E(\tau) = E_0 \, e^{\xi \, \tau}$. If $N_0$ particles
enter the accelerator at $\tau=0$, the number that remains at the time
that corresponds to $\tau$ is: $N(\tau)= N_0 \; e^{-P\, \tau}$.

If all the particle that are injected in the accelerator
have initial energy $E_0$, the spectrum $N_{\rm out} (E)$ of the particles
that exit the source is then:
\begin{equation}
 N_{\rm out} (E) = \left [-\dot{N}(\tau)~\frac{d\tau}{dE} \right ]_{\tau = \tau(E)}
 = \frac{N_0}{E_0} ~(\alpha-1) 
 \; \left ( \frac{E}{E_0} \right )^{-\alpha}
\label{eq:spectrum1}
\end{equation}
where the spectral index is $\alpha = 1+P/\xi$, and 
the normalization satisfies the condition
that integrating over all energies $E \ge E_0$ one obtains $N_0$.

It is commonly accepted that the cosmic ray accelerators are associated
to strong shocks propagating in a magnetized medium.
In this situation the acceleration cycle corresponds to the
motion of a charged particle that diffusing in the medium traverses twice the shock front
(returning to the side where it started).
The constant $\xi$ is then related to the difference between the velocities
of the gas in the upstream and downstream regions:
$\xi \simeq 4/3 \, (u_1-u_2)/c$,
while the escape probability $P$ is related to the
velocity $u_2$ of the shocked gas
in the shock rest frame: $P \simeq 4 \, u_2$.
The velocities $u_1$ and $u_2$ are related by the well
known Rankine--Hugoniot relations,
so that for a mono--atomic gas one has:
$u_1/u_2 = 4/(1 + 3/M^2)$, with $M$ is the shock Mach number $M = u_1/v_s$ with $v_s$ the sound velocity
in the medium. Therefore the spectral index associated to shock acceleration is
\begin{equation}
\alpha = 1+ \frac{P}{\xi} \simeq 2 + \frac{4}{M^2}~.
\label{eq:alpha_fermi}
\end{equation}
In the limit of strong shocks $M \gg 1$ one has that the acceleration spectrum
acquires a universal shape with spetral index $\alpha \to 2$.

The power--law for of the spectrum in Eq.~(\ref{eq:spectrum1}),
is a consequence of the assumption that 
the adimensional quantities $\xi$ and $P$ are
energy (and time) independent.
The result remains valid also when 
$\xi$ and $P$ have a non trivial time (or energy) dependence,
of the same form, so that the ratio $P/\xi$ remains constant.
If this condition is not satisfied the spectrum
generated by the accelerator is not a simple power--law.

As an elementary ``toy model'' one can assume
that $\xi$ is constant, while the probability $P$
has a time (or energy) dependence of the simple form:
\begin{equation}
 P = P_0 + \xi \; b \; \tau = P_0 + b \; \ln (E/E_0) 
\label{eq:prob_b}
\end{equation}
with $b$ a new adimensional parameter.
This equation can only be valid in a finite range
of $\tau$ and $E$ because $P$ can only take value in the interval [0,1],
however for small $b$ the equation
can be considered as a reasonably good approximation in the range
of energy of interest.

The calculation of the accelerator spectrum in
a model where the escape probability 
has the form of Eq.~(\ref{eq:prob_b}) is elementary.
In this case the survival probability of
particles in the accelerator takes the form
$N(\tau)/N_0 = \exp[-(P_0 \, \tau + b \, \tau^2/2)]$
and, following the same steps as before,
one obtains:
\begin{equation}
 N_{\rm out} (E) 
 = \frac{N_0}{E_0} ~[(\alpha -1) + 2 \, \beta \, \ln (E/E_0)] ~
 \; \left ( \frac{E}{E_0} \right )^{-\alpha - \beta \, \ln(E/E_0)}
\label{eq:spectrum2}
\end{equation}
where
\begin{equation}
 \alpha = 1+ \frac{P_0}{\xi}
\end{equation}
and 
\begin{equation}
 \beta =\frac{b}{2 \, \xi} ~.
\end{equation}
The spectrum of Eq.~(\ref{eq:spectrum2}) deviates from the log--parabola
only with logarithmic corrections, with a spectral index
that (for a parameter $b > 0$) grows continuously with energy.
In conclusion, the introduction of Eq.~(\ref{eq:prob_b})
of an escape probability per acceleration cycle that grows with time
(or more in general a ratio $P/\xi$ that grows with time) results
in a curved, gradually softening spectrum that can be reasonably well
approximated with the log--parabola form.

The idea of a ratio $P/\xi$ that grows with time (or equivalently with the
energy of the particles under acceleration) can be realized in several natural ways.
In fact even in the standard Fermi acceleration, the shock slows down
as energy is transfered to shocked gas. This implies that 
the ratio $P/\xi$ increases and the spectrum softens.
Another possible mechanism for a gradual softening of the spectrum
is a modification of the estimate of the escape
probability to take into account the fact the the shock is not a plane,
so that the probability increases for particles of high rigidity.

For other discussions of the formation of log--parabola spectra in
blazar jets see also \cite{Massaro:2003sx,Massaro:2004cc,Massaro:2005qg,Dermer:2015jta}.

\clearpage

\end{document}